\documentclass[twoside]{article}

\usepackage{PRIMEarxiv}

\usepackage[utf8]{inputenc} 
\usepackage[T1]{fontenc}    
\usepackage{hyperref}       
\usepackage{url}            
\usepackage{booktabs}       
\usepackage{amsfonts}       
\usepackage{nicefrac}       
\usepackage{microtype}      
\usepackage{lipsum}
\usepackage{fancyhdr}       
\usepackage{graphicx}       
\graphicspath{{media/}}     
\usepackage{svg}
\usepackage{cancel}
\usepackage{multicol}
\usepackage{hyperref}
\usepackage{booktabs}
\usepackage[utf8]{inputenc}
\usepackage[cmex10]{amsmath}
\usepackage{algorithm}
\usepackage{algorithmic}
\renewcommand{\figurename}{Fig.}
\renewcommand{\tablename}{Table}

\newcommand{\figref}[1]{\textcolor{blue}{\figurename~\ref{#1}}}
\newcommand{\tabref}[1]{\textcolor{blue}{\tablename~\ref{#1}}}

\hypersetup{
    colorlinks=true,
    linkcolor=blue,
    citecolor=blue,
    filecolor=blue,
    urlcolor=blue
}

\usepackage{multirow}

\title{Estimation of Equivalent SCR for Offshore Wind
\thanks{\textit{\underline{\textbf{Accepted} for publication in Electric Power Systems Research and will be presented at Power 
Systems Computational Conference, Cyprus 2026.}} \\} 
}

\author{
Nicolae Darii \\
Siemens Gamesa Renewable Energy A/S \\
Technical University of Denmark \\
Kgs. Lyngby, Denmark \\
\texttt{nidar@dtu.dk}
\And
Ranjan Sharma \\
Siemens Gamesa Renewable Energy A/S \\
Kgs. Lyngby, Denmark \\
\texttt{ranjan.sharma@siemensgamesa.com}
\And
Germano Rugendo Mugambi \\
Technical University of Denmark \\
Kgs. Lyngby, Denmark \\
\texttt{gemuga@dtu.dk}
\And
Oscar Sabor\'{i}o-Romano \\
Technical University of Denmark \\
Kgs. Lyngby, Denmark \\
\texttt{email@email}
\And
Nicolaos A. Cutululis \\
Technical University of Denmark \\
Kgs. Lyngby, Denmark \\
\texttt{email@email}
}

\begin{document}
\maketitle

\begin{abstract}
The integration of offshore wind power plants (OWPPs) into weak grids can pose stability challenges due to the interaction between inverter-based resources (IBRs), Flexible AC Transmission Systems (FACTS) and the grid. In this context, long HVAC transmission systems, relatively common for OWPPs, can exacerbate the stability challenges. Therefore, this paper introduces a novel methodology for estimating the equivalent short-circuit ratio (ESCR) at the offshore point of connection (PoC), combining analytical two-port network (TPN) modeling with electromagnetic transient (EMT) simulations. The approach derives the Thévenin equivalent impedance for passive and active components, enabling accurate ESCR computations without complex derivations. Limitations of traditional SCR metrics are addressed by incorporating the dynamics of the converters, such as static synchronous compensators (STATCOMs), into a hybrid EMT-TPN method for synthesizing equivalent impedances. The process is then verified on the CIGRE OWPP benchmark and is found to capture ESCR variations with cable lengths, shunt reactors, and grid strength. Additionally, the results emphasize the correlation between the ESCR and voltage stability, highlighting the role of STATCOMs in supporting voltage stability in weak grids. This modular framework aids in OWPP design and stability analysis for converter-dominated systems.
\end{abstract}


\section{Introduction}
The rapid growth of renewable energy sources (RES), primarily comprised of inverter-based resources (IBR), is increasing net-zero energy production capacity. However, this growth is also introducing sensitivities to the grid's strength where IBRs are connected \cite{Fan2023AOscillations}. This is due to the strong dependence of phase-locked loops (PLLs) in IBR controls on the grid voltage. This characteristic is unique to grid-following (GFL) inverters, which currently make up the majority of installed IBRs \cite{Zhao2020TransientGrid}. The sensitivity to the grid weakening is particularly evident in offshore wind power plants (OWPPs), where inverter-based (i.e. type-4) wind turbines (WTs) are connected through long HVAC transmission systems, which play a degrading role in terms of overall equivalent impedance at the WTs terminals. \cite{CIGREWorkingGroupB4.622016ConnectionNetworks}. In general, the equivalent impedance is correlated with the voltage sensitivity, as indicated by the short-circuit ratio (SCR) and the equivalent short-circuit ratio (ESCR) \cite{Henderson2024GridSystems}. Since there is a strong correlation between the (E)SCR and the stability limits of IBRs, grid strength information is generally considered during OWPP design for appropriately tuning the WT controls. Having accurate data on the (E)SCR is therefore essential for ensuring stable operation \cite{Fu2020CriticalGrid}. 
The SCR at the onshore point of connection (PoC) is typically provided by the transmission system operators (TSOs). However, the provided index may differ significantly from the one at the medium-voltage terminals of WTs, i.e. at the end of the offshore transmission system. The mismatch arises from the influence of passive (e.g., reactors, transformers, cables) and active components (e.g., STATCOMs and nearby OWPPs) connected to the offshore transmission system and the WT terminals. As a result, offshore WTs (with controls tuned using the SCR given by the TSO) may not perform as expected, potentially impacting stability. It is therefore necessary to translate the SCR given at the onshore PoC into an equivalent SCR (ESCR) at the offshore PoC to evaluate site-specific grid integration issues effectively. The proposed method addresses this need by enabling the identification of the correct ESCR at the offshore PoC based on the one provided by the TSO at the onshore PoC, or by allowing TSOs to directly identify and provide it themselves at the offshore PoC.

Stability and design studies require the screening of grid conditions at the point where the OWPP is to be connected \cite{Mugambi2025MethodologiesAnalysis}. Typically, it is not possible to represent the grid in detail unless the TSO performs the studies.
The TSO thus provides information about the grid conditions at the onshore PoC in the form of a Thévenin equivalent. However, this approach is insufficient for studies involving OWPPs, as the actual WT PoC is far from that of the onshore grid, and a complex transmission system lies in between.    

Relatively simple analytical ways to represent the offshore transmission network employing series RL components \cite{Guerreiro2025PassiveVerification}, or considering fewer shunt components such as a single static VAR compensator (SVC) \cite{Ghimire2024ImpactPlants} can be found in the literature.
Either way, the formulation does not account for the complexity of offshore transmission cables or other elements such as filter capacitors. In addition, the formulation becomes even more complex when even a few components are added \cite{CIGREWorkingGroupB4.622016ConnectionNetworks}. In attempting to overcome such challenges, this work proposes the use of two-port networks (TPNs) \cite{Benato2010EHVPlanning}, which enables more accurate modeling of complex components such as long cables through simple matrix multiplications.

When active converter-based components are also involved, the purely analytical estimation of the Thévenin equivalent becomes even more challenging \cite{Boricic2022SystemGrids}. The challenge is exacerbated by the lack of access to the converters' internal control structures due to intellectual property (IP) protection. In the traditional numerical EMT-simulation-based method, the equivalent impedance is estimated directly from the measured short-circuit current. This approach is limited to estimating absolute values of the equivalent impedance and does not guarantee an accurate estimation e.g. if converters reach their limits during the process.

This work introduces a methodology combining analytical techniques based on two-port network (TPN) modeling, with numerical electromagnetic transient (EMT) simulations to estimate the grid
strength at the offshore PoC, henceforth referred to as the equivalent short-circuit ratio (ESCR).
Unlike previous studies that 
derive closed-form expressions for specific network configurations \cite{CIGREWorkingGroupB4.622016ConnectionNetworks,Ghimire2024ImpactPlants}, this work aims at a general approach that can be applied to any offshore transmission system topology. The proposed technique offers additional features: modularity, given by the properties of the TPN matrices, and the ability to extract resistive and inductive elements of the ESCR separately, thanks to its complex nature.
Moreover, it presents an EMT-based ESCR analysis tool that computes ESCR values directly from time-domain simulations.
This eliminates the need for lengthy analytical derivations and overcomes the inaccessibility of specific converter internal control structures when models are provided in black-box form. Additionally, the numerical extracted version can be implemented as a TPN and lately used as a module for the overall calculation of the ESCR.

Each method is first introduced, followed by the mathematical derivations that relate one to the other. 

A comprehensive analysis then considers scenarios where a single method may be insufficient, thus requiring a mixed analytical-numerical approach.

The rest of the paper is organized as follows. 

Section \ref{Methods} pertains to the derivation of the mixed analytical–EMT approach  
using TPN modeling and Thévenin equivalent estimation to evaluate the Equivalent Short Circuit Ratio (ESCR) in offshore grids with inverter-based resources. 
The method is verified in Section \ref{results} through EMT simulations on the CIGRE benchmark, with parametric studies on cables, shunt reactors, network strength, and STATCOM integration, showing strong alignment between analytical and EMT methods. 
The simulation results are discussed in Section \ref{discussion}, and concluding remarks are given
in Section \ref{conclusion}.

\section{Methodology}
\label{Methods}
\subsection{Thévenin Equivalent Impedance of a Two-Port Network}

\begin{figure}
    \centering
    \includegraphics[width=1\linewidth]{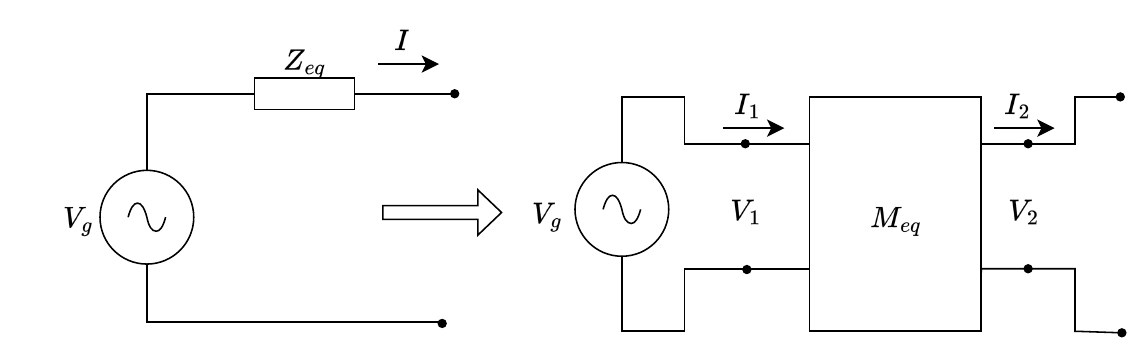}
    \caption{Two-port network Thévenin equivalent}
    \label{fig:double_b_thevein}
\end{figure}

The basic version of the Thévenin equivalent, comprised of an ideal voltage source with an equivalent RL impedance and proposed in \cite{LANDON1930TheTheorem}, can be converted into a TPN version as shown in \figref{fig:double_b_thevein}. The voltage $V_{\text{g}}$ of the equivalent ideal voltage source is equal to the no-load (i.e. $I_2=I_1=I=0$) voltage at the $V_2$ terminals of the TPN. The equivalent impedance $Z_{\text{eq}}$ can be found by substituting the system's voltage sources with short circuits.
The principle is then applied to the system modeled as an equivalent TPN with the corresponding transmission matrix $\mathbf{M}_{\text{eq}}$ \eqref{eqn: double_b}, whose elements are complex numbers identifying the relationships between the input-output current-voltage and differ depending on the device, as shown in \tabref{tab:transmission_m}.

\begin{equation}
    \begin{bmatrix}
\overline{V}_1 \vphantom{\displaystyle\sum} \\
\overline{I}_1
\end{bmatrix}
= \mathbf{M}_{\text{eq}}
\begin{bmatrix}
\overline{V}_2 \vphantom{\displaystyle\sum} \\
\overline{I}_2
\end{bmatrix} =
\begin{bmatrix}
\overline{A}_{\text{eq}} \vphantom{\displaystyle\sum} & \overline{B}_{\text{eq}} \\
\overline{C}_{\text{eq}} & \overline{D}_{\text{eq}}
\end{bmatrix}
\begin{bmatrix}
\overline{V}_2 \vphantom{\displaystyle\sum} \\
\overline{I}_2
\end{bmatrix}
\label{eqn: double_b}
\end{equation}

If the Thévenin equivalent definition is applied and the input voltage $\overline{V}_1$ is replaced with a short circuit, it is possible to expand the formulation \eqref{eq: th_b_1} and find the equivalent impedance at terminals $V_2$ \eqref{eq:th_b_2}:

\begin{subequations}
\begin{align}
0 = \overline{A}_{\text{eq}} \cdot \overline{V}_2 + \overline{B}_{\text{eq}} \cdot \overline{I}_2 \label{eq: th_b_1} \\
 \overline{Z}_{\text{eq}} = \frac{\overline{V}_2}{\overline{I}_2} = -\frac{\overline{B}_{\text{eq}}}{\overline{A}_{\text{eq}}} \label{eq:th_b_2}
\end{align}
\end{subequations}

\subsection{Estimation through transmission matrices}

\begin{figure}
    \centering
    \includegraphics[width=1\linewidth]{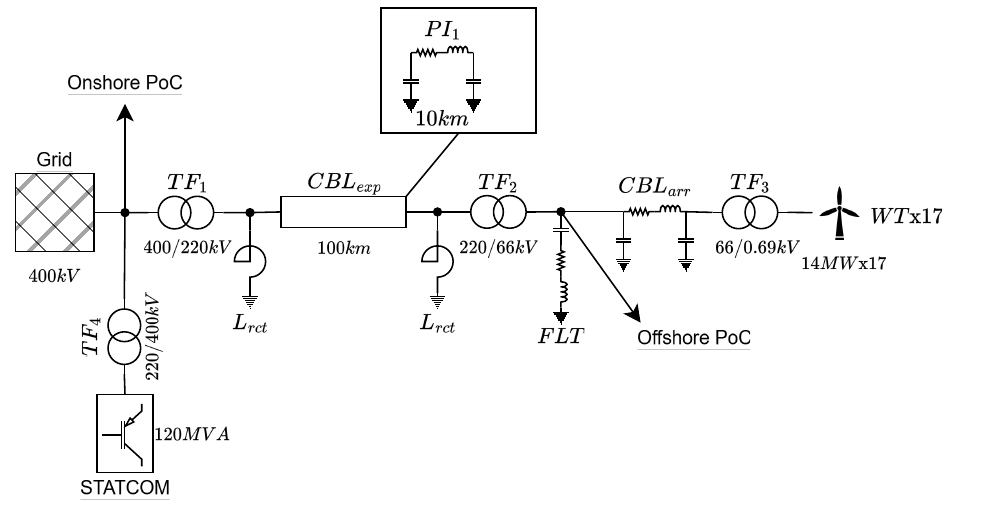}
    \caption{Offshore wind power plant benchmark}
    \label{fig:OWPP_Scheme}
\end{figure}

Two-port networks have a straightforward application on an OWPP that is entirely comprised of passive components. Therefore, the passive components of the OWPP benchmark, shown in \figref{fig:OWPP_Scheme}, can be modeled as TPNs via the transmission matrices given in \tabref{tab:transmission_m}.

\begin{table}
    \centering
    \small
    \caption{Transmission matrix of passive OWPP components}
    \renewcommand{\arraystretch}{1.3}
    \begin{tabular}{|l|l|}
        \hline
        \textbf{Element} & \textbf{Parameter and Matrix} \\
        \hline
        Grid impedance & 
        \( \vphantom{\begin{bmatrix} 1 \\ 1 \\ 1 \end{bmatrix}} \mathbf{M}_{\text{g}} = \begin{bmatrix} 1 & \bar{Z}_{\text{g}} \\ 0 & 1 \end{bmatrix} \) \\
        \hline
        Transformer & 
        \( m = {V_{\text{high}}}/{V_{\text{low}}} \) \\
         & 
        \( \vphantom{\begin{bmatrix} 1 \\ 1 \\ 1 \end{bmatrix}} \mathbf{M}_\text{TF} = \begin{bmatrix} m & {\bar{Z}_\text{cc}}/{m} \\ 0 & {1}/{m} \end{bmatrix} \) \\
        \hline
        Shunt Reactor / Filter & 
        \( \vphantom{\begin{bmatrix} 1 \\ 1 \\ 1 \end{bmatrix}} \mathbf{M}_{r/f} = \begin{bmatrix} 1 & 0 \\ \left({R + jX_\text{L} - jX_\text{C}}\right)^{-1} & 1 \end{bmatrix} \) \\
        \hline
        Cable & 
        \( \bar{z} = r + j\omega l, \quad \bar{y} = g + j\omega c \) \\
         &
        \( \bar{k} = \sqrt{\bar{z} \cdot \bar{y}}, \quad
           \bar{Z}_{\text{c}} = \sqrt{{\bar{z}}/{\bar{y}}} \) \\
         & 
        \( \vphantom{\begin{bmatrix} 1 \\ 1 \\ 1 \end{bmatrix}} \mathbf{M}_{\text{c}} = \begin{bmatrix}
        \cosh(\bar{k}L) & \bar{Z}_{\text{c}} \sinh(\bar{k}L) \\
        \sinh(\bar{k}L)/{\bar{Z}_{\text{c}}} & \cosh(\bar{k}L)
        \end{bmatrix} \) \\
        \hline
    \end{tabular}
    \label{tab:transmission_m}
\end{table}
The multiplication property of the transmission matrices enables the derivation of an equivalent transmission matrix representing all passive components in the OWPP at the offshore PoC \eqref{eqn: multiplication_prop}.

\begin{equation}
    \mathbf{M}_{\text{eq}} = \mathbf{M}_{\text{g}}\cdot \mathbf{M}_{\text{TF1}}\cdot \mathbf{M}_{\text{r}}\cdot \mathbf{M}_{\text{c}}\cdot \mathbf{M}_{\text{r}}\cdot \mathbf{M}_{\text{TF2}}\cdot \mathbf{M}_{\text{f}}
    \label{eqn: multiplication_prop}
\end{equation}

$\mathbf{M}_{\text{eq}}$ is also a transmission matrix with complex elements, to which the TPN Thévenin equivalent \eqref{eq:th_b_2} can be applied. This formulation can potentially account for $n$ parallel export cables or transformers by dividing the distributed resistance, $r$, and distributed inductance, $l$, by $n$, and multiplying the distributed conductance, $g$, and distributed capacitance, $c$, by $n$. In contrast to single-phase one-port formulations  \cite{Ghimire2024ImpactPlants}, the introduction of additional shunt or series components does not produce complex analytical expressions. 

The equivalent impedance, $Z_{\text{eq}}$, is extracted from \eqref{eqn: multiplication_prop} via \eqref{eq:th_b_2} and is then used for the equivalent ESCR at the offshore PoC through \eqref{eqn: ESCR}, which is the reciprocal of the per-unit equivalent impedance at the studied node \cite{CIGREWorkingGroupB4.622016ConnectionNetworks}.

\begin{equation}
    \text{ESCR} = \frac{V_{\text{b}}^2}{S_\text{n}|\overline{Z}_{\text{eq}}|} =\frac{Z_{\text{b}}}{|\overline{Z}_{\text{eq}}|} = \frac{1}{|\overline{Z}_{\text{eq}}|_{\text{pu}}}
    \label{eqn: ESCR}
\end{equation}

$V_{\text{b}}$ is the nominal voltage at the offshore PoC (e.g. $66\,\text{kV}_\text{RMS,L-L}$), and $S_{\text{n}}$ is the nominal power of the OWPP.
In addition, since all the components of the transmission matrix (including shunt capacitances) are present and the $\mathbf{M}_{\text{eq}}$ has complex elements, it is possible to synthesize the
equivalent inductance, $L_{\text{eq}}$, and equivalent resistance, $R_{\text{eq}}$, at the offshore PoC by simply extracting it from $\overline{Z}_{\text{eq}}$ \eqref{eqn: LRextract}. The transmission matrix thus allows a clearer characterization of the ESCR’s resistive and inductive components.

\begin{subequations}
\begin{align}
R_{\text{eq}} =  \Re(\overline{Z}_{\text{eq}})\\
L_{\text{eq}} = \frac{\Im(\overline{Z}_{\text{eq}})}{2\pi f_{\text{nom}}} 
\end{align}
\label{eqn: LRextract}
\end{subequations}

\subsection{Estimation through EMT simulation}
\begin{figure*}
    \centering
    \includegraphics[width=1\linewidth]{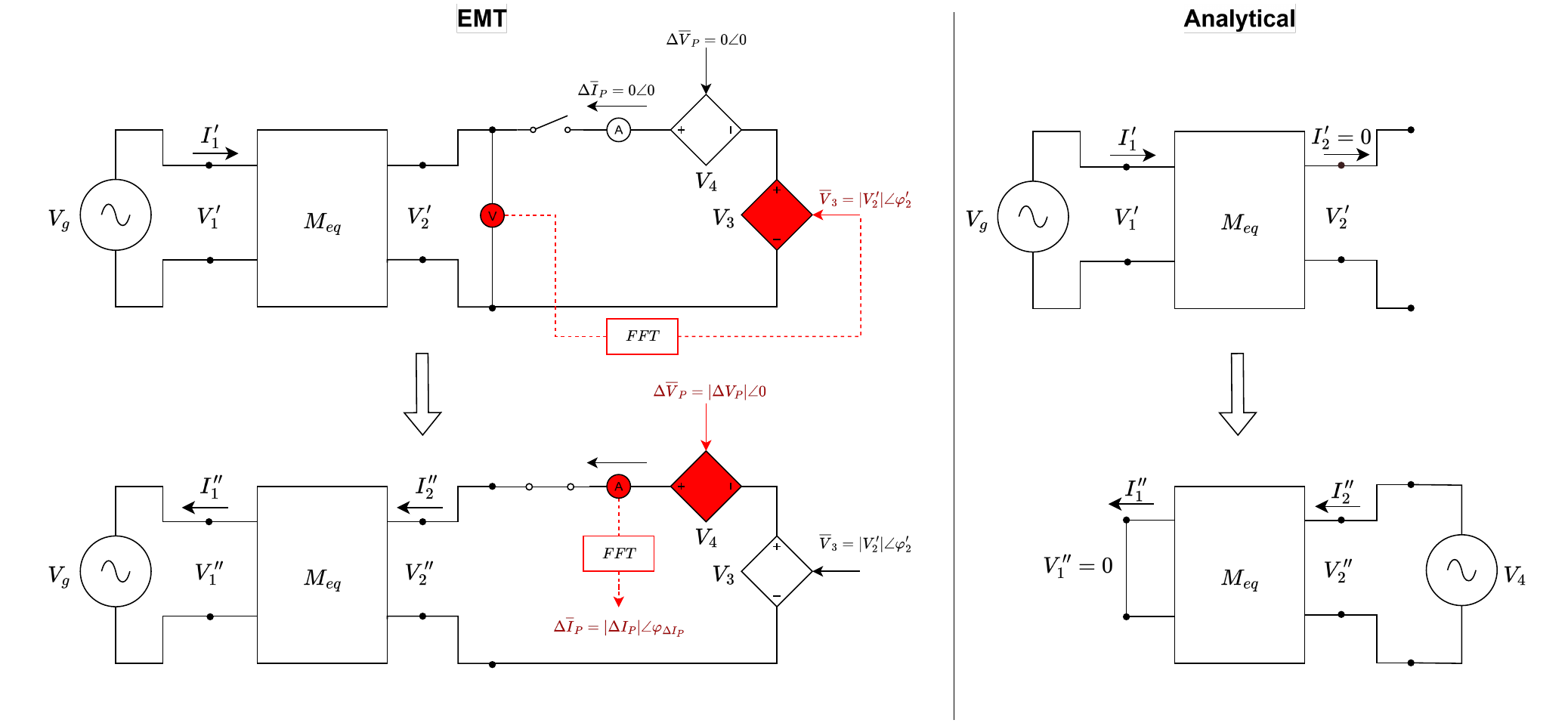}
    \caption{ESCR estimation through EMT simulation}
    \label{fig:ESCRinEMT}
\end{figure*}

The numerical estimation of the ESCR through EMT simulation is worth analyzing, as it can replace or be combined with analytical methods. 
Since converter-based systems do not lend themselves to the analytical computation of the Thévenin equivalent impedance via TPNs, they need to be adapted for numerical computation through EMT simulation. 
This can be achieved by introducing two controlled voltage sources (CVSs), as illustrated on the left side of \figref{fig:ESCRinEMT}. The process consists of two phases: extraction of the open-circuit values and perturbation of the system.

\subsubsection{EMT-simulation equivalent of open-circuit voltage}
\label{sec:open}

In the first phase, highlighted in red at the top left of \figref{fig:ESCRinEMT}, the complex open-circuit voltage at the $V_2$ terminals of the TPN, $\overline{V}_2'$, is derived from \eqref{eqn: openM} and given by \eqref{eqn: openM1}, which corresponds to the Ferranti Effect in complex form derived via the transmission matrix \cite{Grigsby2017PowerStability}. 
Variables estimated in this phase are denoted with a prime ($^\prime$).

\begin{equation}
\begin{bmatrix}
\overline{V}_1' \vphantom{\displaystyle\sum} \\
\overline{I}_1'
\end{bmatrix}
=
\begin{bmatrix}
\overline{A}_{\text{eq}} \vphantom{\displaystyle\sum} & \overline{B}_{\text{eq}} \\
\overline{C}_{\text{eq}} & \overline{D}_{\text{eq}}
\end{bmatrix}
\begin{bmatrix}
\overline{V}_2' \vphantom{\displaystyle\sum} \\
0
\end{bmatrix}
\label{eqn: openM}
\end{equation}
\begin{equation}
    \overline{V}_1' = \overline{A}_{\text{eq}} \cdot \overline{V}_2' \Rightarrow \overline{V}_2' = \frac{\overline{V}_{1}'}{\overline{A}_{\text{eq}} } = \frac{\overline{V}_{\text{g}}}{\overline{A}_{\text{eq}} }
    \label{eqn: openM1}
\end{equation}

The same principle can be applied to EMT simulation by introducing two CVSs, which are disconnected from the system during the initial analysis.  
The instantaneous open-circuit line-to-ground voltage is first measured at the $V_2$ terminals. 
The magnitude, $|\overline{V}_2'|=V_2'$, and phase, $\varphi_2'$, are then extracted from the measurements through an online fast Fourier transform (FFT). Once it reaches steady state, $\overline{V}_{2}'$ is recorded and assigned to the CVS $\overline{V}_3$. Such voltage must be equal to the one derived analytically \eqref{eqn: openM1}.

If carried out correctly, the procedure should yield the following results.
\begin{itemize}
    \item No current will flow through the CVSs, $\overline{I}_2'=0$, as expressed in \eqref{eqn: openM}.
    \item A current $\overline{I}_1'$ may still flow into the shunt components (if present), as given by \eqref{eqn: I_void}.
    \begin{equation}
        \overline{I}_1' = \overline{C}_{\text{eq}} \cdot \overline{V}_2' = \frac{\overline{C}_{\text{eq}}}{\overline{A}_{\text{eq}} }\overline{V}_{1}' = \frac{\overline{C}_{\text{eq}}}{\overline{A}_{\text{eq}} }\overline{V}_{\text{g}}
        \label{eqn: I_void}
    \end{equation}
    \item If shunt elements or converters are present, the voltage assigned to $\overline{V}_3$ will be different than the upstream voltage, $\overline{V}_{\text{g}}$, as given in \eqref{eqn: openM1}. Prior knowledge of $\overline{V}_{\text{g}}$ is not required.
\end{itemize}

\subsubsection{EMT-simulation equivalent of voltage source substitution}

In the second phase, highlighted in red at the bottom left of \figref{fig:ESCRinEMT}, \eqref{eqn: double_b} becomes 
\eqref{eqn: closedM}. Variables estimated in this phase are denoted with a double prime ($^{\prime\prime}$). It is straightforward to impose $\overline{V}_1''=0$ analytically, which is equivalent to substituting the voltage sources with short circuits.
However, such substitution is not possible in EMT simulation of black-box systems or systems with converters, which is why the two CVSs are introduced. 
The first CVS, $\overline{V}_3$, is set to the steady-state open-circuit voltage $\overline{V}_2'$ defined in the first phase.
Consequently, a known voltage perturbation $\Delta \overline{V}_{\text{P}}$ is applied to the second CVS, $\overline{V}_4$, preferably with zero phase. This is equivalent to short-circuiting the voltage sources in the system, $\overline{V}_1=0$, as expressed in \eqref{eqn: closedM}, shown in \figref{fig:ESCRinEMT}, and demonstrated in the following.

\begin{equation}
\begin{bmatrix}
0 \\
\overline{I}_1'' \vphantom{\displaystyle\sum} 
\end{bmatrix}
=
\begin{bmatrix}
\overline{A}_{\text{eq}} \vphantom{\displaystyle\sum} & \overline{B}_{\text{eq}} \\
\overline{C}_{\text{eq}} & \overline{D}_{\text{eq}}
\end{bmatrix}
\begin{bmatrix}
\overline{V}_2'' \vphantom{\displaystyle\sum} \\
\overline{I}_2''
\end{bmatrix}
\label{eqn: closedM}
\end{equation}

\begin{enumerate}
\item 
Firstly, \eqref{eqn: double_b1} can be derived from \eqref{eqn: double_b} for the short-circuit case by denoting the variables with a double prime ($''$). It should be noted that in this case $\overline{V}_1''\neq 0$ since it may not be possible to replace the voltage sources with short circuits in EMT simulation.

\begin{equation}
\overline{V}_1'' = \overline{A}_{\text{eq}} \cdot \overline{V}_2'' + \overline{B}_{\text{eq}} \cdot \overline{I}_2''
\label{eqn: double_b1}
\end{equation}

$\overline{V}_1''$ is given by \eqref{eqn: source_cc} for the system shown in \figref{fig:ESCRinEMT}.

\begin{equation}
    \overline{V}_1'' = \overline{V}_{\text{g}}
\label{eqn: source_cc}
\end{equation}

In the equivalent analytical case, shown on the right side of \figref{fig:ESCRinEMT}, $\overline{V}_{\text{g}}$ is the only voltage source behind the equivalent TPN. However, the system may comprise multiple voltage/current sources in different topologies. This does not impact the calculation of the equivalent impedance, and this property can be utilized to synthesize an equivalent converter's impedance, as explained in Section \ref{sec:passive}.

\item 
The voltage applied in EMT simulation at the $V_2$ terminals thus corresponds to the sum of the voltages from the two CVSs, $\overline{V}_{3}, \overline{V}_{4}$, as expressed in \eqref{eqn: step2} and shown at the bottom left of \figref{fig:ESCRinEMT}.

\begin{equation}
    \overline{V}_2'' = \overline{V}_{3} + \overline{V}_{4}
\label{eqn: step2}
\end{equation}

\item 
Set to the steady-state open-circuit voltage $\overline{V}_2'$ defined in the first phase (Section \ref{sec:open}) as \eqref{eqn: openM1}, the voltage of the first CVS, $\overline{V}_3$, can be expressed as in \eqref{eqn: step3}.

\begin{equation}
    \overline{V}_{3} = \overline{V}_2' = \frac{\overline{V}_{1}'}{\overline{A}_{\text{eq}} } = \frac{\overline{V}_{\text{g}}}{\overline{A}_{\text{eq}}}
\label{eqn: step3}
\end{equation}

\item The second CSV, $\overline{V}_{4}$, takes the value of the known voltage perturbation, $\Delta\overline{V}_{\text{P}}$, which produces a current $\overline{I}_2''=\Delta \overline{I}_{\text{P}}$. 

\item Consequently, \eqref{eqn: source_cc}, \eqref{eqn: step2} and \eqref{eqn: step3} can be substituted in \eqref{eqn: double_b1}, thus producing \eqref{eqn: sobstitution}, in which there are no upstream variables. 

\begin{equation}
    \cancel{\overline{V}_{\text{g}}} = \cancel{\overline{A}_{\text{eq}} \cdot \frac{\overline{V}_{\text{g}}}{\overline{A}_{\text{eq}}}} + \overline{A}_{\text{eq}} \cdot \Delta\overline{V}_{\text{P}}+\overline{B}_{\text{eq}} \cdot \Delta 
    \overline{I}_{\text{P}}
\label{eqn: sobstitution}
\end{equation}

\item 
Therefore, by simplifying \eqref{eqn: sobstitution} and comparing it with \eqref{eq: th_b_1}, the proposed EMT simulation adaptation is coherent with the analytical method, as expressed in \eqref{eq: realtion}
\begin{equation}
0 = \overline{A}_{\text{eq}} \cdot \overline{V}_2 + \overline{B}_{\text{eq}} \cdot \overline{I}_2 \Leftrightarrow  0 = \overline{A}_{\text{eq}} \cdot \Delta\overline{V}_{\text{P}} + \overline{B}_{\text{eq}} \cdot \Delta \overline{I}_{\text{P}}
\label{eq: realtion} 
\end{equation}

\end{enumerate}

In conclusion, although it is not possible to directly replace all the voltage sources with short circuits as theoretically required in \cite{LANDON1930TheTheorem}, it is still possible to replicate the analytical procedure by correctly assigning the voltages to two CVSs and waiting for the EMT simulation to reach steady state.

Since the EMT simulation process relies on online FFT, it is intrinsically affected by aliasing, which depends on the sampling frequency, thus simulation timestep 
\cite{On-LineScanning}. Therefore, the simulation timestep must be selected in accordance with the Nyquist–Shannon sampling theorem applied to the system being studied.  The process with the TPN is therefore consistent with the process using simple series components proposed in \cite{Boricic2022SystemGrids}.

Finally, the ratio between the perturbed voltage and current is computed from \eqref{eq: realtion}. The result is the ratio between two complex elements of the system's transmission matrix, which, as can be seen from \eqref{eq:th_b_2}, is the equivalent impedance of the system \eqref{eqn: EMT_impedance}, $\overline{Z}_{\text{eq}}$.

\begin{equation}
    \frac{\Delta V_{\text{P}} \angle 0}{\Delta I_{\text{P}} \angle \varphi_{\Delta I_{\text{P}}} } = -\frac{\overline{B}_{\text{eq}}}{\overline{A}_{\text{eq}}} =\overline{Z}_{\text{eq}}
    \label{eqn: EMT_impedance}
\end{equation}

\subsection{Hybrid Approach} \label{sec:mixed}

If the EMT-simulation-based approach cannot be directly applied (e.g. when converter-based active components such as a STATCOM are connected at the onshore PoC, as shown in \figref{fig:OWPP_Scheme}), a hybrid 
approach can be employed instead, as presented in the following.

\begin{figure}
    \centering
    \includegraphics[width=0.8\linewidth]{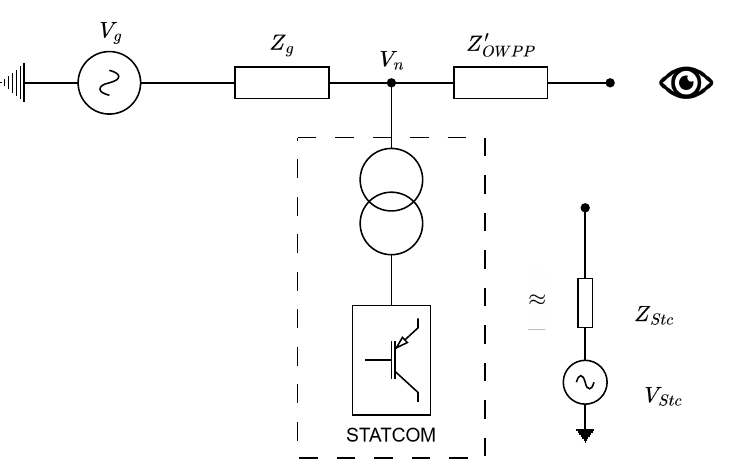}
    \caption{Simplified offshore wind power plant representation}
    \label{fig:OWPPEquivalent}
\end{figure}

ESCR estimation should include the case of an open circuit at the OWPP end, but a STATCOM may malfunction due to incompatibility between its control mode/gains and the open-circuit topology.
It is thus worth exploring alternatives that are compatible with the open-circuit topology, as the study must be conducted with the STATCOM operating normally and the OWPP connected. 
Moreover, the overall system can be simplified by representing the OWPP complex passive network with an equivalent RL impedance and the STATCOM shunt as a Thévenin equivalent, as proposed in \cite{Moharana2014SSRLine} and shown in \figref{fig:OWPPEquivalent}.

At a high level, the process is as follows.
\begin{enumerate}
    \item EMT-simulation-based estimation is performed with the active components/devices, checking if the converter-based devices remain in stable operation. In this case, there is no need to simplify the OWPP network (\ref{par1}).
    \item If active converter-based devices do not remain in stable operation, a simplified version of the OWPP network is obtained.
    \item The ESCR of the simplified OWPP is estimated through EMT simulation with the converter-based devices connected (\ref{sec:passive}).
    \item Finally, an equivalent shunt reactance is calculated for each converter-based device and used in the original network, both in EMT simulation and analytically (\ref{par3}).
\end{enumerate}

\subsubsection{OWPP simplification}
\label{par1}
The first step involves simplifying the passive components of the OWPP by analytically extracting the equivalent impedance $\overline{Z}_{\text{OWPP}}$ from \eqref{eqn: equivalent OWPP} via \eqref{eq:th_b_2}. Then, the equivalent impedance is translated to the high-voltage side of $TF_1$, becoming $\overline{Z}_{\text{OWPP}}'$.

\begin{equation}
    \mathbf{M}_{\text{OWPP}} = \mathbf{M}_{\text{TF1}}\cdot \mathbf{M}_{\text{r}}\cdot \mathbf{M}_{\text{c}}\cdot \mathbf{M}_{\text{r}}\cdot \mathbf{M}_{\text{TF2}}\cdot \mathbf{M}_{\text{f}}
    \label{eqn: equivalent OWPP}
\end{equation}

The grid impedance, $\overline{Z}_{\text{g}}$, either is known by design or can be 
estimated by this method
at the onshore PoC if the network is provided. 

Additionally, only one voltage source is assumed upstream of the equivalent TPN in \eqref{eqn: source_cc}. For the simplified OWPP in  \figref{fig:OWPPEquivalent}, it is clear that the open-circuit voltage measured downstream, $\overline{V}_{\text{n}}'$, is a weighted average based on the generators’ voltages and the system impedance \eqref{eqn: simplifiedOWPPopen}, as given by Millman's theorem \cite{Aguirre-Zamalloa2009MillmansRevisited}.

\begin{equation}
\overline{V}_{\text{n}}' = \frac{\dfrac{\overline{V}_{\text{g}}}{\overline{Z}_{\text{g}}} + \dfrac{\overline{V}_{\text{Stc}}}{\overline{Z}_{\text{Stc}}}}{\dfrac{1}{\overline{Z}_{\text{g}}} + \dfrac{1}{\overline{Z}_{\text{Stc}}}}
\label{eqn: simplifiedOWPPopen}
\end{equation}

The EMT-simulation-based estimation is then performed with two voltage sources, where the first takes the value of the open-circuit voltage $\overline{V}_{\text{n}}'$, and the second takes the value of a known voltage perturbation, $\Delta\overline{V}_{\text{P}}$. Thus, the voltage applied in EMT simulation at bus $V_n$, $\overline{V}_{\text{n}}''$, is given by \eqref{eqn: simplifiedOWPPopen1}, and the current produced by the voltage perturbation, $\Delta\overline{I}_{\text{P}}$, is given by \eqref{eqn: simplifiedOWPPopen3}.

\begin{equation}
\overline{V}_{\text{n}}'' = \frac{\dfrac{\overline{V}_{\text{g}}}{\overline{Z}_{\text{g}}} + \dfrac{\overline{V}_{\text{Stc}}}{\overline{Z}_{\text{Stc}}}+ \dfrac{\overline{V}_{\text{n}}' + \Delta\overline{V}_{\text{P}}}{\overline{Z}_{\text{OWPP}}'}}{\dfrac{1}{\overline{Z}_{\text{g}}} + \dfrac{1}{\overline{Z}_{\text{Stc}}}+\dfrac{1}{\overline{Z}_{\text{OWPP}}'}}
\label{eqn: simplifiedOWPPopen1}
\end{equation}

\begin{equation}
    \Delta\overline{I}_{\text{P}} = \dfrac{\Delta\overline{V}_{\text{P}} + \overline{V}_{\text{n}}' - \overline{V}_{\text{n}}''}{\overline{Z}_{\text{OWPP}}'}
    \label{eqn: eqn: simplifiedOWPPopen2}
\end{equation}

The equivalent impedance (considering only the variables accessible from the downstream terminal) is then given by \eqref{eqn: simplifiedOWPPopen3}. 
This impedance corresponds to that which would be found find if the classical method was applied and all voltage sources were replaced.

\begin{equation}
    \dfrac{\Delta\overline{V}_{\text{P}}}{\Delta\overline{I}_{\text{P}}}  = \frac{\overline{Z}_{\text{g}} \overline{Z}_{\text{Stc}} + \overline{Z}_{\text{g}} \overline{Z}_{\text{OWPP}}'+ \overline{Z}_{\text{Stc}} \overline{Z}_{\text{OWPP}}'}{\overline{Z}_{\text{g}} +\overline{Z}_{\text{Stc}}}
    \label{eqn: simplifiedOWPPopen3}
\end{equation}
The methodology holds as long as its premise remains valid, i.e. the system elements can be equivalently modeled as voltage/current sources or shunt elements.

\subsubsection{Synthesis of equivalent passive components}
\label{sec:passive}
To estimate the STATCOM's equivalent impedance, $\overline{Z}_{\text{Stc}}$, the EMT-simulation-based method is first applied to estimate the impedance at the WT/WPP end of the simplified network in \figref{fig:OWPPEquivalent}, $\overline{Z}_{\text{f}}$, for the desired operating point. 
Calculating $\overline{Z}_{\text{f}}$ as the corresponding Thévenin equivalent impedance \cite{LANDON1930TheTheorem}, i.e. with every voltage source in the simplified network replaced with a short circuit, results in \eqref{eqn: z_parallel}.

\begin{equation}
    \overline{Z}_{\text{f}}  = \frac{\overline{Z}_{\text{eq}} \overline{Z}_{\text{Stc}}}{\overline{Z}_{\text{eq}} +\overline{Z}_{\text{Stc}}} + \overline{Z}_{\text{OWPP}}'
    \label{eqn: z_parallel}
\end{equation}

\begin{figure}
    \centering
    \includegraphics[width=0.8\linewidth]{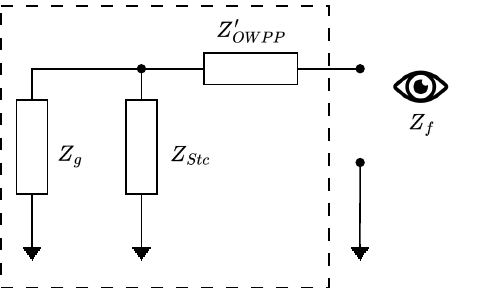}
    \caption{Equivalent final impedance}
    \label{fig:OWPPEquivalentZ}
\end{figure}

Therefore, only the Thévenin equivalent impedance of the STATCOM shunt in \figref{fig:OWPPEquivalent}, $\overline{Z}_{\text{Stc}}$, is needed for estimating the ESCR \cite{Moharana2014SSRLine}, 

as shown in \figref{fig:OWPPEquivalentZ}.
In particular, 
 
$\overline{Z}_{\text{Stc}}$ is determined not only by the STATCOM's filter and transformer impedances, but also by its converter output impedance, which is shaped by its internal controls.

\subsubsection{ESCR estimation with equivalent passive component}
\label{par3}
Finally, an equivalent shunt reactor (or similarly: capacitor), $L_{\text{Stc}}$, can be calculated \eqref{eqn: Lstat}.

\begin{equation}
    L_{\text{Stc}} =
    \dfrac{\Im\left[\dfrac{\overline{Z}_{\text{eq}}(\overline{Z}_{\text{OWPP}}'-\overline{Z}_{\text{f}})}{\overline{Z}_{\text{eq}}+\overline{Z}_{\text{OWPP}}'-
    \overline{Z}_{\text{f}}}\right]}{2\pi f_{\text{nom}}}
    \label{eqn: Lstat}
\end{equation}

The equivalent shunt reactor matrix form (given in \tabref{tab:transmission_m}), $\mathbf{M}_{\text{Stc}}$, is computed analytically \eqref{eqn: LstatAnalyt} and then used to represent the STATCOM in EMT simulation. Hence, it is possible to recompute the ESCR at the offshore PoC using either the EMT-simulation-based or analytical approaches.

\begin{multline}
\mathbf{M}_{\text{eq}} = \mathbf{M}_{\text{g}} 
\cdot 
\mathbf{M}_{\text{Stc}} 
\cdot 
\mathbf{M}_{\text{TF1}} 
\cdot 
\mathbf{M}_{\text{r}} 
\cdot 
\mathbf{M}_{\text{c}} 
\cdot 
\mathbf{M}_{\text{r}} 
\cdot 
\mathbf{M}_{\text{TF2}} 
\cdot 
\mathbf{M}_{\text{f}}\ ,\\
\mathbf{M}_{\text{Stc}} = \begin{bmatrix} 1 & 0 \\[6pt] \dfrac{1}{j 2 \pi L_{\text{Stc}}} & 1 \end{bmatrix}
\label{eqn: LstatAnalyt}
\end{multline}

The estimation of the equivalent inductance $L_{\text{Stc}}$ with \eqref{eqn: Lstat} should be repeated whenever the grid impedance changes, and a look-up table, $L_{\text{Stc}} = f({\text{SCR}}_{\text{Grid}}, {\text{X/R}}_{\text{Grid}})$, should be composed, since the active effect of the STATCOM would result in different control efforts depending on the grid impedance.

\section{Results}
\label{results}
This section aims to demonstrate the equivalence between the analytical modeling of the OWPP elements and the EMT extraction method. Furthermore, it will illustrate the design opportunities enabled through analytical TPN modeling of the system and ESCR extraction through an EMT-based method.
To illustrate the method's capabilities, the CIGRE OWPP benchmark  \cite{CIGREWorkingGroupC4.492024Multi-frequencySystems} will be used. Specifically, a $240 \text {MW}$ OWPP with an HVAC connection of $100 \text{km}$ with shunt compensators (displayed in \figref{fig:OWPP_Scheme}),  will be utilized to conduct different types of studies:

\begin{itemize}
    \item Different HVAC cable lengths
    \item With and without shunt reactors
    \item Influence of upstream network change in strength
    \item STATCOM during different grid strengths
\end{itemize}

Subsequently, the results of the converter-based system extraction will be presented, with the EMT extraction method as the primary technique. The analytical approach will serve as both a verification method and a supplementary tool to assist in the extraction of the converter's equivalent impedance (in this specific case of a STATCOM).

Therefore, the equivalent values of the ESCR of the offshore PoC with and without STATCOM are compared to the response generated by a detailed switching original equipment manufacturer (OEM) model of an offshore wind turbine. In this scenario, the EMT extraction method will be the primary approach, with the analytical method serving as both a verification and a supplementary technique to facilitate extraction in converter-based systems. 

Lastly, the ESCR values will be juxtaposed with the voltage and reactive power response of the detailed switching OEM model of an offshore wind turbine, to verify if the voltage sensitivity through the ESCR has actual feedback on the WT's reaction.
The results will support the three principal research objectives:

\begin{itemize}
    \item Describe how the TPN-based analytical method and the EMT one are functioning and are equivalent.
    \item Demonstrate how it is possible to model the equivalent impedance more accurately, when OWPP is composed of long HVAC cables and multiple shunt elements. In addition, how can this affect the OWPP design.
    \item Show how the mixed method can aid the estimation of systems of converters. Additionally, showing how the estimated index is reflected in the actual OWPP WTs reactions.
\end{itemize}

\subsection{Passive case}
\begin{figure}
    \centering
    \includegraphics[width=0.8\linewidth]{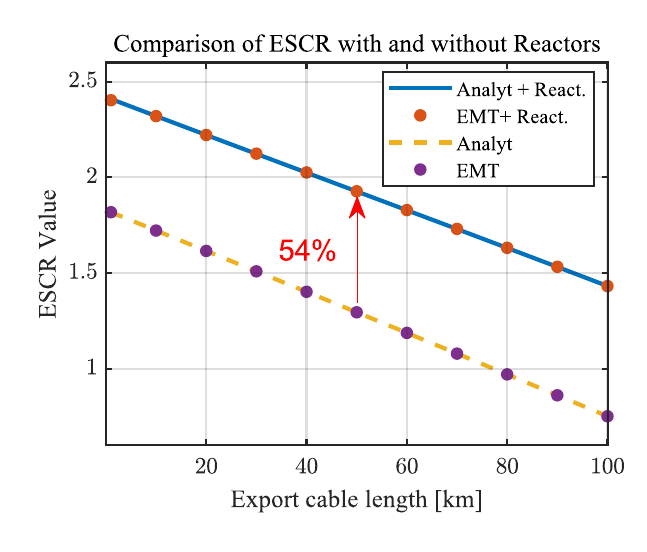}
    \caption{ESCR dependent on calbe's length}
    \label{fig: calb-len}
\end{figure}

The first application involves testing how the length of the HVAC cable affects the ESCR at the offshore PoC. This allows for parametric studies. The ESCR identified through TPNs is then verified via EMT simulation on discrete measurements, as shown in \figref{fig: calb-len}, showing that the estimated values match for both methods. Additionally, the impact of shunt reactors used to compensate for the high capacitive effect of the HVAC is also taken into account. In this case, using the TPN, any HVAC passive components can be easily connected through matrix multiplications, as demonstrated in \eqref{eqn: equivalent OWPP}. Consequently, the offshore PoC ESCR is compared when shunt reactors and filters are present in the OWPP. It is evident that the presence of shunt components, primarily inductive, adds an extra benefit to ESCR, alongside reactive power compensation. More specifically, the two main shunt reactors, $L_{\text{rct}}$, shown in 
\figref{fig:OWPP_Scheme}, are designed to compensate for 70\% of the total cables' capacitive effect. This prevents overcompensation and minimizes voltage drop caused by series inductive and resistive losses during the regular operation of the WTs. In the benchmark case, these shunt reactors are shown to improve the ESCR by 54\%, and the EMT discrete measurements the analytical results for both cases.

\begin{figure}
    \centering
    \includegraphics[width=0.8\linewidth]{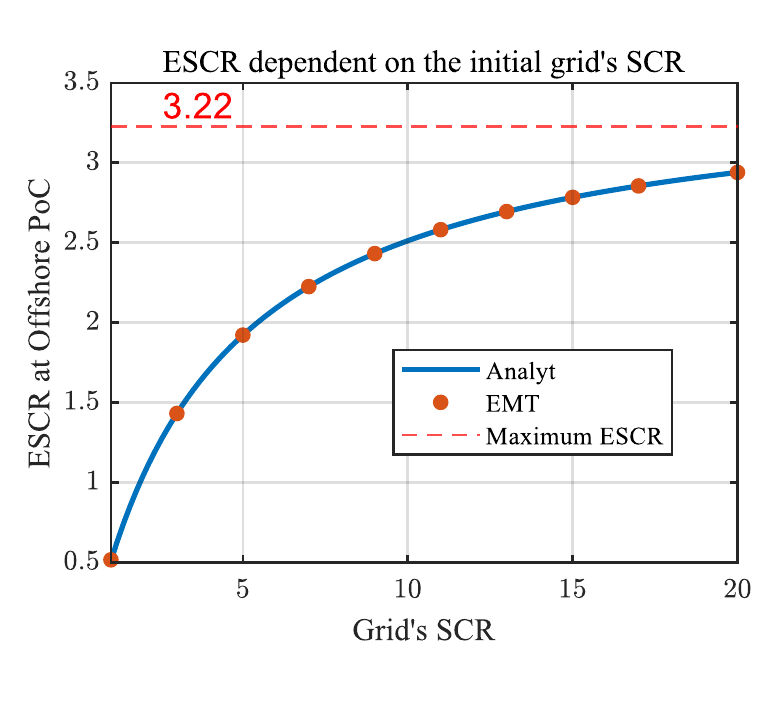}
    \caption{ESCR dependent on the initial grid's SCR}
    \label{fig: SCR-change}
\end{figure}

In the benchmark case, the equivalent model of the OWPP upstream grid is employed as a Thevenin equivalent. A parametric analysis was  performed, as depicted in \figref{fig: SCR-change}, demonstrating how the offshore ESCR changes when the upstream network modifies its strength.
The ESCR does not vary linearly; it declines more rapidly at
low grid SCRs, while it tends asymptotically to a value of $3.22$
under conditions of a very high grid SCR.
The analytical results were corroborated through comparison with EMT simulations.

\begin{figure}
    \centering    \includegraphics[width=0.8\linewidth]{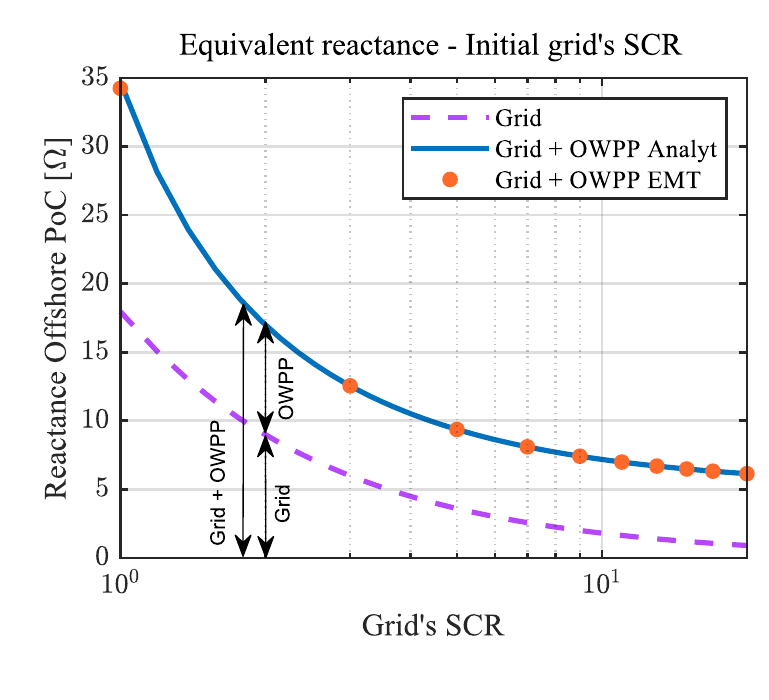}
    \caption{Equivalent reactance of grid and OWPP at 66 kV}
    \label{fig: X-change}
\end{figure}

Additionally, through TPN, it is possible to decompose the equivalent impedance at the $66kV$ level of the OWPP and identify it. Additionally, when the upstream network has a significant SCR, the OWPP dominates the equivalent impedance at the offshore PoC. Therefore, the equivalent reactance extracted by EMT matches the analytical value, visible in \figref{fig: X-change}.

\subsection{Active case}

The second application results estimate the offshore ESCR when a converter, such as a STATCOM, is connected at the onshore, as shown in \figref{fig:OWPP_Scheme}. 

\begin{figure}
    \centering
    \includegraphics[width=0.8\linewidth]{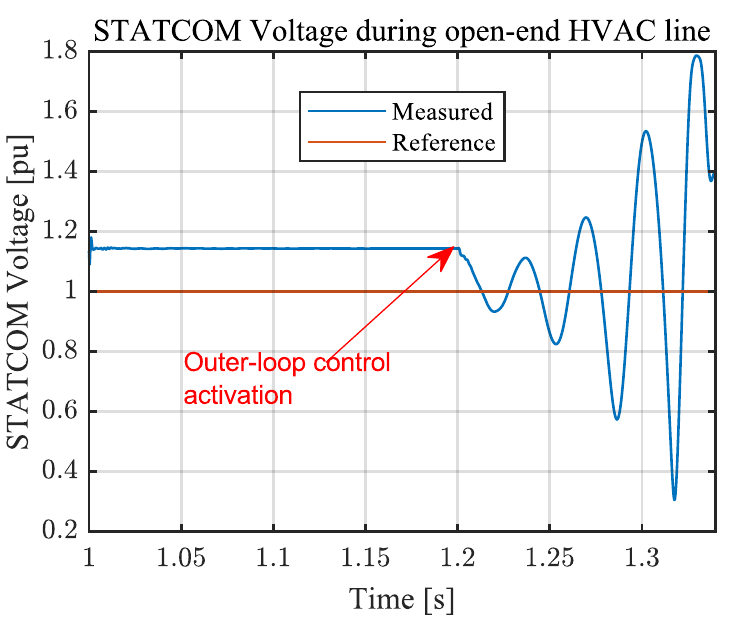}
    \caption{STATCOM with open-end HVAC line}
    \label{fig: unstable}
\end{figure}

Generally, if the converter-based system allows the direct ESCR extraction in EMT, no additional steps are needed. However, in this setup, the STATCOM is designed to operate only when WTs are connected at the offshore level; the specific FACTS cannot function and become asymptotically unstable once the outer-loop voltage controller is activated, as shown in  \figref{fig: unstable}.

For this reason, it was necessary to employ the mixed analytical-EMT method explained in \ref{sec:mixed}. Therefore, the equivalent RL series impedance was extracted in order to redefine the system in a simplified version as shown in \figref{fig:OWPPEquivalent}. From the Analytical extraction (confirmed then through EMT) the equivalent $Z_{\text{OWPP}}'$ resulted in an equivalent resistance $R_{\text{OWPP}}'$ of $18.1920 \ \Omega$ and an equivalent inductance $L_{\text{OWPP}}'$ of $0.6004 \ H$. 

The STATCOM was connected to the simplified system, and several sensitivity tests were performed (and given in \tabref{tab: ops}) to verify whether the extracted equivalent impedance would result in a different equivalent impedance when responding to different levels of disturbances:

\begin{itemize}
    \item Different ESCR estimator voltage perturbations;
    \item Change the initial grid's voltage;
    \item Change the STATCOM voltage set-point;
\end{itemize}

Despite having different reactive power injected at each operational point, the EMT-based impedance and ESCR remain unchanged, while the X/R ratio exhibits a minor divergence. 

\begin{figure}
    \centering
    \includegraphics[width=0.8\linewidth]{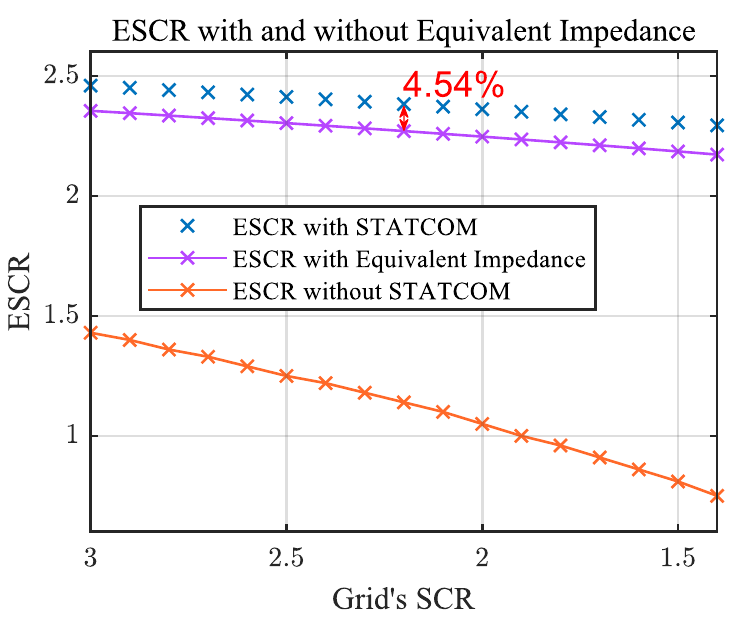}
    \caption{ESCR with and without Equivalent Impedance}
    \label{fig: equivalent_stat}
\end{figure}

\begin{table}
    \centering
    \caption{ESCR sensitivity during different operational points}
    \begin{tabular}{ccccccc}
        \toprule 
        $V_{\text{p}}$ & $V_{\text{g}}$ & $V_{\text{Stc}}$ & $Q_{\text{Stc}}$ & \multirow{2}{*}{${\text{ESCR}}$} & $L_{\text{f}}$ & \multirow{2}{*}{$X/R_{\text{f}}$}\\  
        
        [\%] & [pu] & [pu] & [pu] & & [$H$] \\
        \hline 
        1 & 1 & 1 & -0.0313 & 2.46 & 0.86 & 13.92 \\
        5 & 1 & 1 & -0.159 & 2.46 & 0.86 & 13.86 \\
        1 & 0.9 & 1 & +0.340 & 2.46 & 0.86 & 14.90 \\
        1 & 1.1 & 1 & -0.430 & 2.46 & 0.86 & 13.26 \\
        1 & 1 & 0.9 & -0.389 & 2.46 & 0.86 & 13.20 \\
        1 & 1 & 1.1 & 0.244 & 2.46 & 0.86 & 14.72 \\
        \bottomrule
    \end{tabular}
    
    \label{tab: ops}
\end{table}
Consequently, the equivalent ESCR ad X/R ratio from the \tabref{tab: ops} was used to extract $Z_{\text{f}}$ through \eqref{eqn: EMT_impedance}, the equivalent impedance of the OWPP $Z_{\text{OWPP}}'$ and the one known from the grid $Z_{\text{g}}'$. Finally, the equivalent inductance $L_{\text{Stc}}$ is computed with \eqref{eqn: Lstat} and, in this case, it is equal to $0.4072 H$. 
The tests given in \tabref{tab: differen_ESCRs} were performed for SCR ranges from 3 to 1.4, and the results are shown in \figref{fig: equivalent_stat}.

\begin{table}
    \centering
    \caption{Tested Cases}
    \begin{tabular}{ccccc}
    \toprule
        Device & Setup & EMT & Anaytical\\
        \midrule
       STATCOM  & \figref{fig:OWPPEquivalent} & x &  \\
        Equivalent Inductance $L_{\text{Stc}}$ & \figref{fig:OWPP_Scheme} & x & x \\
        None & \figref{fig:OWPP_Scheme} & x & x \\
        \bottomrule
    \end{tabular}
    
    \label{tab: differen_ESCRs}
\end{table}

It is possible to notice that there is a sliglty mismatch of $4.54\%$ from the ESCRs computed with the STATCOM in the simplified case compared to the ones computed with the equivalent inductance $L_{\text{Stc}}$ in the whole system, but following the exact pattern while the grid's SCR degrades.

\subsection{Benchmark}

\begin{figure}
    \centering
    \includegraphics[width=0.8\linewidth]{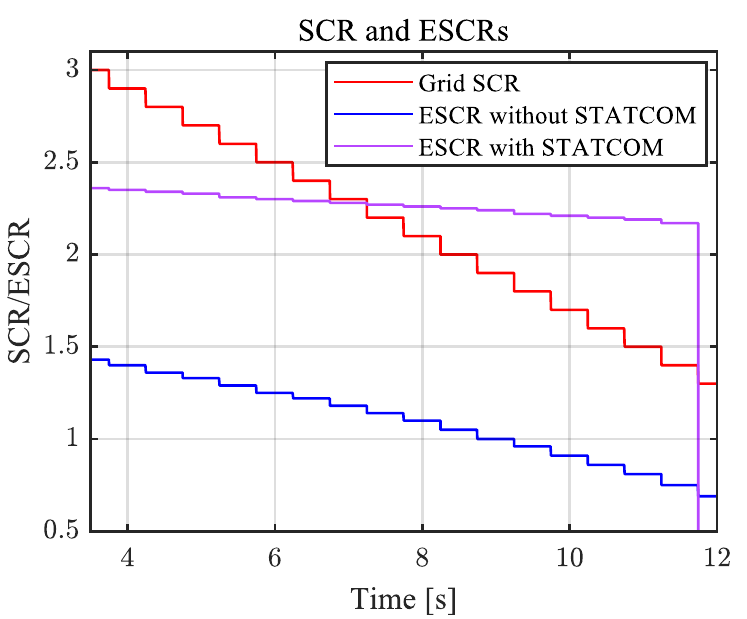}
    \caption{ESCR and SCR at Grid and WT level}
    \label{fig: full_ESCR_SCR}
\end{figure}

The last test was done to correlate the ESCR change (with and without STATCOM) with the reaction of the offshore WTs. 
To perform the test, the grid's steady state was set with a variable SCR that starts at $3$ and decreases by $0.1$ every $ 0.5$ seconds, allowing for the oscillations to stabilize before the subsequent SCR decrease. The decreasing SCR pattern is shown on \figref{fig: full_ESCR_SCR}, where  the ESCR (computed previously offline) is also shown, highlighting the impact of the STATCOM.

The ESCR at the offshore level varies significantly, especially with low grid SCR, regardless of STATCOM. When SCR drops below 2.3, the offshore PoC's ESCR exceeds that of the grid with STATCOM. This may differ with IBR converters that inject active power.

In addition, once the grid's SCR reaches 1.4, the ESCR seen at the offshore level drops sharply. This happens because the grid resistance and inductance at $400kV$ are respectively $57 \ \Omega$ and $1.6 \ H$. Which, if translated at $220kV$, becomes $20 \Omega$ and $0.58 H$ (Including the transformer $TF_4$). 

\begin{figure}
    \centering
    \includegraphics[width=0.8\linewidth]{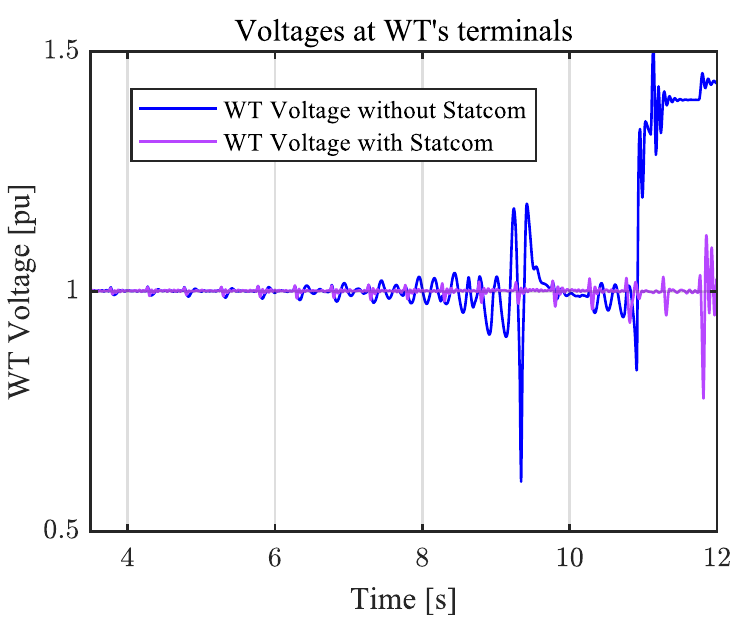}
    \caption{WT's voltage with and without STATCOM}
    \label{fig: WTs-Volt}
\end{figure}

This behaviour is also reflected in the WT's terminal voltage, as shown in \figref{fig: WTs-Volt}.  With the STATCOM present, the voltage remains stable and highly damped until the grid's SCR reaches the STATCOM's stability limit. On the contrary, when the STATCOM is not present, the offshore PoC results in way lower ESCR, resulting in highly undamped voltage oscillations after the grid's SCR equal to 2.3. 

\begin{figure}
    \centering
    \includegraphics[width=0.8\linewidth]{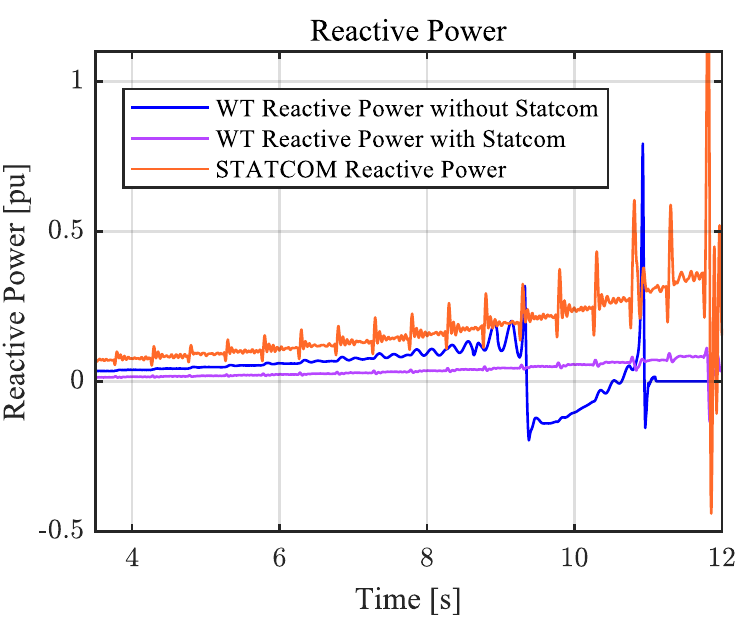}
    \caption{Reactive power of WT and STATCOM}
    \label{fig: reactive}
\end{figure}

It is helpful to look at the reactive power, where it is possible to see in \figref{fig: reactive} that without the STATCOM, the WTs, which are far from the grid, try to compensate for the voltage oscillations, but without a significant effect. On the other hand, when the STATCOM is connected, it reacts relatively fast to the voltage deviations by absorbing reactive power when necessary.

\section{Discussion}
\label{discussion}



\subsection{Literature Context}
The methodology proposed in this work differs from traditional analyses at the offshore wind power plant (OWPP) level, such as those presented in \cite{CIGREWorkingGroupB4.622016ConnectionNetworks}. The Two-Port Network (TPN) approach enables a lighter and more modular formulation, which is not achievable with lumped single-phase components that require complete reformulation for any structural change, as highlighted in \cite{Ghimire2024ImpactPlants}. Moreover, the TPN accurately captures the upstream system’s non-ideal inertial characteristics, evidenced by the fact that the current $I_1$ is not necessarily zero, and by the representation of downstream voltage variations (Ferranti Effect) through the related equations.

\subsubsection{Modeling of Equivalent Impedance}
The extraction of the Thévenin equivalent impedance results in a complex impedance value representing both resistive and inductive parts. This differs from conventional methods based solely on the magnitude of short-circuit power or the nominal power of actual short-circuit current contributors \cite{Damanik2022EvaluationSystems, Chen2016EffectiveSTATCOM}. Consequently, the proposed method provides a more accurate representation of the equivalent upstream impedance seen at the offshore Point of Connection (PoC).

\subsubsection{Representation and Role of converters}
In converter-dominated systems, which inherently lack high short-circuit currents \cite{Bennett2023GridSystem, Zhang2014EvaluatingIntegration}, conventional metrics predict a reduced system strength. However, studies such as \cite{Li2016ImpactSCR, Narula2024EmpoweringCapabilities} show that converters like STATCOMs reduce voltage oscillations, contradicting the notion of degraded strength when evaluated solely by short-circuit criteria. This work addresses this contradiction by analyzing the active impact of converters and their control schemes on the perceived equivalent impedance at the offshore PoC during regular operation—not only during faults, when converters saturate as noted in \cite{Boricic2022SystemGrids}. The EMT estimation performed on a TPN-derived simplified system enables assessment of converters' contributions dynamically. Therefore, defining the equivalent inductive element representing the converter yields the same ESCR as the original system, including the converter.

\subsection{Methodological Contributions}
From an analytical standpoint, the TPN modeling offers an accurate and modular way to represent offshore equivalent impedance while clarifying the underlying principles of the EMT-based ESCR estimation—thus providing theoretical grounding beyond pure empirical approaches. The EMT method further allows software-based ESCR computation with a firm theoretical correspondence. The proposed hybrid analytical-EMT technique enables non-invasive ESCR estimation in systems with converters, overcoming limitations of traditional short-circuit-based methods.

\subsection{Application and Insights from the Benchmark Case}
Applying the methodology to the CIGRE offshore benchmark with an onshore STATCOM reveals that despite injecting reactive power, the STATCOM behaves effectively as an equivalent inductance, corroborating its operation similar to a synchronous condenser rather than an SVC. The converter’s control results in a relatively constant equivalent impedance, while the amplitude of the controlled voltage source manages voltage regulation. For HVAC OWPP design, the modular approach facilitates parametric studies of the wind turbine (WT) stability limits depending on various OWPP components—serving as an additional decision-making index when considering conversion to HVDC structures or the adequacy of adding STATCOMs or SVCs to meet minimum ESCR requirements at the offshore PoC.

\subsection{Extensions and Future Work}

The proposed method opens the possibility for further research and development, including:
\begin{itemize}
    \item Correlating the nominal ratings (power, voltage, control gains, etc.) of converters with their equivalent shunt inductance to enable prediction of contributions without direct measurement.
    \item Relating the impedance matrix obtained via frequency scanning to the equivalent transmission matrix of the system.
    \item Extending the method to frequencies other than nominal, especially higher frequencies where transmission matrix modeling becomes crucial due to traveling wave effects.
    \item Expanding the hybrid approach for isolating equivalent impedances of internal converter controllers.
\end{itemize}

\section{Conclusion}
\label{conclusion}
This study presents a methodology for estimating the Equivalent Short Circuit Ratio (ESCR) in offshore wind farms. It combines the modular approach of the TPN analytical modeling with the non-intrusive nature of EMT-based simulations to analyze black-box systems. By accurately considering the influence of passive transmission components and active elements like STATCOMs, this method addresses the limitations of traditional SCR evaluations, providing deeper insights into system strength and stability. Verification with the CIGRE benchmark demonstrates the method's consistency across different techniques and its ability to accurately predict the converters impedance. This framework supports parametric design of offshore wind farms—such as cable length, shunt compensation, and converters contributions—and sets the stage for future improvements, including frequency-dependent analyses and predictive converter modeling. Ultimately, it ensures reliable estimation of grid strength, enabling proper tuning or retuning of offshore turbines and assessing the feasibility of offshore wind projects.

\section{Acknowledgements}
This work is supported by the European Union as part of ADOreD project funded by the Horizon Europe MSCA programme (\href{https://www.msca-adored.eu/}{HORIZON-MSCA-2021-DN, Grant agreement 101073554})

\section{Legal Disclaimer}
Figures and values presented in this paper should not be used to judge the performance of Siemens Gamesa Renewable Energy technology as they are solely presented for demonstration purpose. Any opinions or analysis contained in this paper are the opinions of the authors and not necessarily the same as those of Siemens Gamesa Renewable Energy.

\bibliographystyle{unsrt}  
\bibliography{references}

\end{document}